\documentclass[preprint,12pt]{elsarticle}

\usepackage{amssymb}
\usepackage{amsmath}
\usepackage{subfiles}
\usepackage{hyperref}

\begin{document}

\begin{frontmatter}



\title{PyPAS - Python package for Positron Annihilation Spectroscopy Doppler Broadening Analysis} 


\author[aff_1,aff_2]{Achiya Yosef Amrusi}[orcid=0009-0004-5512-9722]\cormark[1]
\author[aff_2]{Sharon May-Tal Beck}[orcid=0000-0002-0409-4222]
\author[aff_1]{Hadar Steinberg}[orcid=0000-0002-7409-5087]
\author[aff_1]{Guy Ron}[orcid=0000-0002-8129-7146]

\cortext[1]{Corresponding author. Email: \href{mailto:ahia.amrosi@mail.huji.ac.il}{ahia.amrosi@mail.huji.ac.il}}


\affiliation[aff_1]{organization={Racah Institute of Physics, The Hebrew University of Jerusalem},%
                     city={Jerusalem},%
                     postcode={91904},%
                     country={Israel}}
\affiliation[aff_2]{organization={Physics department, NRCN},%
                     city={Beer-Sheva 84190},%
                     postcode={9001},%
                     country={Israel}}

\begin{abstract}
Doppler Broadening (DB) of  annihilation radiation is a well-established technique within Positron Annihilation Spectroscopy (PAS), used for probing the electronic structure of  materials. The analysis of DB experimental data relies on gamma spectroscopy analysis tools, while depth profiling using variable-energy slow positron beams depends on solving the positron diffusion equation. Traditional Variable Energy Doppler Broadening (VEDB) analysis tools, such as VEPFIT and ROYPROF, often present limitations due to outdated interfaces and lack of integration with comprehensive spectroscopy analysis platforms. Addressing these challenges, an open-source Python package for PAS analysis, PyPAS, is introduced. PyPAS offers functionalities including Coincidence Doppler Broadening (CDB) filtering, two-dimensional CDB analysis with DB and resolution extraction, and computation of lineshape parameters (S and W). Furthermore, it integrates modules for generating thermal positron implantation profiles based on established models, solving positron diffusion equations using finite-difference methods and optimizing diffusion length. This work presents the architecture of the PyPAS package and the validation results and demonstrates the application of the package through case studies.
\end{abstract}

\begin{graphicalabstract}
\end{graphicalabstract}

\begin{highlights}
\item Enables modern analysis of Doppler broadening PAS data.
\item Includes modules for Coincidence Doppler broadening filtering and lineshape parameter extraction.
\item Implements a finite-difference solver for positron diffusion equations.
\item Extracts implantation profiles using Makhovian and Ghosh models.
\item Validated against analytical and SciPy-based numerical solutions.
\item Provides open-source, modular design for reproducible PAS workflows.
\item Challenges previous two-layer models (40 nm and bulk) suggested by the VEPFIT code using AIC.
\end{highlights}


\end{frontmatter}



\section{Introduction}\label{Introduction}

Positron Annihilation Spectroscopy (PAS) \cite{Tuomisto2013,Selim2021} is a well-established field that encompasses various experimental techniques to study the electronic structure of materials.
Specifically, in the Doppler Broadening (DB) method, the energy shift of the electron-positron annihilation photons is measured. Because the kinetic energy of the positron is usually low compared to the energy of the electrons because of the positron thermalization, the energy and angle Doppler shift correspond to the electron momentum and thus are used to study the electronic structure.
Slow positron beams are commonly used to control the positron energy which dictates the penetration depth distribution. Thus, they allow for the characterization of materials ‘layer by layer’ from the surface to the bulk. Slow positron beams provide approximately monoenergetic positrons, commonly in a range of 0 - 50 [keV], enabling the use of Variable Energy Doppler Broadening (VEDB) methods. In VEDB, the sample is scanned at multiple beam energies, enabling the extraction of intrinsic material parameters, such as the positron diffusion length as a function of depth.  
Analysis of VEDB measurement of a sample requires a solver for the positron diffusion equation, calculation of the thermal positron implantation profile, and optimization algorithm to extract the optimal diffusion lengths for a set of measurements. There are several analysis codes available for VEDEB, such as VEPFIT, ROYPROF, and e+DSc \cite{Veen1991, Saleh1999, Dryzek2021}. However, most of the programs used traditionally in the field are old, are not open source, some require specific computer operation system, and use a GUI which limits the user workflow. Moreover, these codes are not integrated with spectroscopy tools and other analysis methods.
\texttt{PyPAS} is a modern open-source Python package based on the \texttt{PySpectrum} spectroscopy analysis code to handle and analyze PAS Doppler Broadening measurements. 
\texttt{PyPAS} methods include the Coincidence Doppler Broadening (CDB) filter, 2D CDB analysis with DB and resolution extraction, as well as the calculation of the line-shape parameters (S and W parameters) of the measured peaks. \texttt{PyPAS} contains additional modules for the generation of thermal positron implantation profiles based on Makhovian and Ghosh methods \cite{Ghosh1995, Dryzek2008} as well as a numerical solver for the positron diffusion equation. The diffusion solver is based on the finite-differences method and includes the electric-field term. Lastly, \texttt{PyPAS} has a material and layered-sample description module as well as an optimization tool module for various optimization such as diffusion length extraction. This enable the use of the \texttt{PyPAS} for VEDB analysis within a Python environment. 
The code is uploaded to GitHub and is available at- \url{https://github.com/achiyaAmrusi/pyPAS}. The code is well documented and includes several jupyter-notebook examples with simulated and real data, to aid new users.
This paper presents a detailed description together with the validation of the pyPAS solver and verification of the code with diffusion length optimization. 

\clearpage

\section{Background}\label{Background}
\subsection{Positron Interaction with Matter}
\begin{figure}
\centering
\includegraphics[width=1\linewidth]{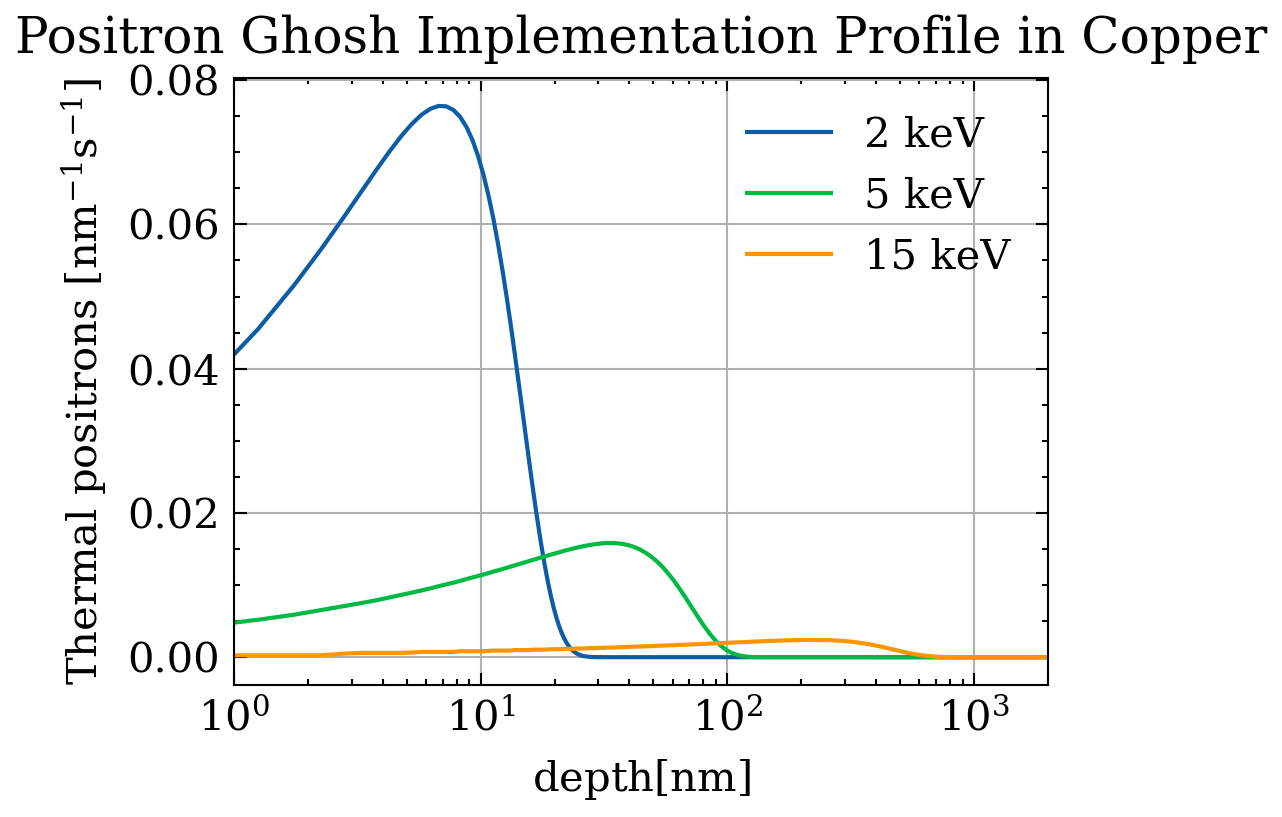}
\caption{The positrons implantation profile in Cu sample according to the Ghosh profile function. The parameters of the profile were taken from reference \cite{Dryzek2008}} \label{fig:ghosh_profiles} 
\end{figure}

Energetic positrons in a material lose their energy rapidly in a series of ionization interactions until thermalization. The thermal positrons then diffuse in the material until annihilation occurs, or they can encounter and get trapped at minimum potential locations in the lattice, such as vacant point defects (e.g. dislocations,vacancies, voids, and gas bubbles) and then annihilate. In the positron-electron annihilation process, two photons are emitted  back-to-back, each carrying 511 keV in the center of mass frame of reference \cite{Selim2021, Tuomisto2013}. Due to the stochastic nature of the ionization-scattering interactions, mono-energetic positrons that enter into a sample become thermal at various depths in the sample. The thermal positron implantation profile is traditionally approximated by the Makhov or Ghosh profiles, where profile parameters depend on both the density and composition of the material and the initial energy of the positrons (Figure~\ref{fig:ghosh_profiles}) \cite{Dryzek2008, Ghosh1995}.

The profile of annihilation events in the material is modeled using the one-dimensional differential equation~\ref{eq:diffusion_equation} with boundary conditions presented in equation~\ref{eq:diffusion_equation_boundary} \cite{Veen1991}. This equation accounts for the implantation, diffusion, drift, and annihilation processes in the bulk and various defect types. Note that the parameter (z) in the equation is the depth in the sample which by convention is parallel to the beam direction.
Based on equation~\ref{eq:diffusion_equation}, the diffusion length of positrons in the material, $L_+$, is defined in equation~\ref{eq:mean_free_path}.
The diffusion equation can be solved using the finite difference method. This method is already implemented in several existing codes \cite{Veen1991, Saleh1999}.

\begin{equation}
D_+ \frac{d^2 c(z)}{dz^2} -  \frac{d}{dz}(v_d(z) \cdot c(z))-  \left(\lambda_b + \sum_d \lambda_d\right) \cdot c(z) = - I(z)
\label{eq:diffusion_equation}
\end{equation}

\begin{equation}
\begin{cases}
\left. \dfrac{dc}{dz} - \dfrac{c}{L_a} \right|_{z=0} = 0, \\
\lim\limits_{z \to \infty} c(z) = 0
\end{cases}
\label{eq:diffusion_equation_boundary}
\end{equation}

Where $c(z)$ is the positron concentration inside the sample; $v_d$ is the drift velocity, which is equal to $ \mu(z) \cdot E(z)$ where $\mu(z)$ is the positron mobility and $E(z)$ is the electric field in the sample;  $L_a$  is the absorption length at the surface (note that this parameter does not affect the line-shape analysis as mentioned in \cite{Veen1991}),   $\lambda_b$ and $\lambda_d$ are the annihilation rates in the bulk and the different types of defects respectively, and finally $I(z)$ is the implantation profile of thermal positrons in the sample.

\begin{equation}
\begin{aligned}
L_+ &= \sqrt{\frac{D_+}{\lambda_b + \sum_d \lambda_d}}
\end{aligned}
\label{eq:mean_free_path}
\end{equation}

\subsection*{Doppler Broadening Spectroscopy (DBS)}
\begin{figure}
\centering
\includegraphics[width=0.95\linewidth]{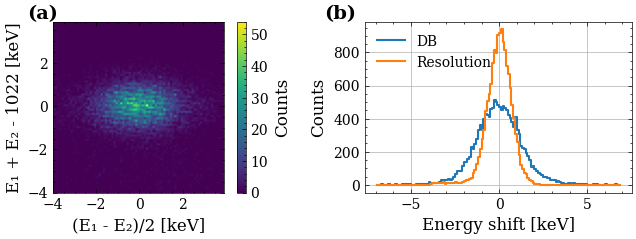}
\caption{\textbf{(a)} Two dimensional CDB (Coincidence Doppler Broadening) histogram measured using the SPOT-IL beam at 5 keV on a copper sample. The horizontal axis corresponds to the Doppler energy shift, while the vertical axis reflects the detector’s energy resolution. \textbf{(b)} The horizontal (blue) and vertical (orange) projections of the 2D CDB histogram, corresponding to the Doppler broadening (DB) and resolution spectra, respectively. The DB spectrum clearly extends beyond the instrument resolution, indicating the presence of physical broadening.}
\label{fig:cdb_2d_example} 
\end{figure}

In DBS, the broadened energy distribution of the annihilation photons is measured using a high resolution spectrometer. The peak is then analyzed to study the electronic structure of the material. Measuring the two annihilation photons using the Coincidence Doppler Broadening (CDB) method allows the measurement of DB with a reduced background and thus higher sensitivity to the electronic structure of the bounded electron \cite{Selim2021}. In CDB the spectrometer is comprised of two face-to-face detectors (or a few such detector couples \cite{BUTTERLING20112623}). An annihilation event is defined by the simultaneous detection of photons in the two detectors, and the requirement that the energy sum of the two photons is 1022 keV, twice the electron rest mass. 
The widths of the distributions of the time difference between the simultaneously measured photons and their energy sum are determined by the combined time resolution of the detectors and their combined energy resolution, respectively. 
The difference between the detected energies of the two annihilation photons is equal to twice the Doppler shift. 
Figure~\ref{fig:cdb_2d_example} presents CDB data from a copper sample measured using the SPOT-IL beam at 5 keV\cite{Or2024}. Panel (a) presents a two-dimensional histogram showing the number of events as a function of the Doppler shift (energy difference) and the detector combined energy resolutions (energy sum), and panel (b) presents horizontal (blue) and vertical (orange) projections of the 2D CDB histogram, corresponding to the DB and resolution spectra, respectively. The DB spectrum is clearly wider than the detectors energy resolution, indicating the presence of physical broadening. 

\subsection{Lineshape Parameters}

\begin{figure}
\centering
\includegraphics[width=0.95\linewidth]{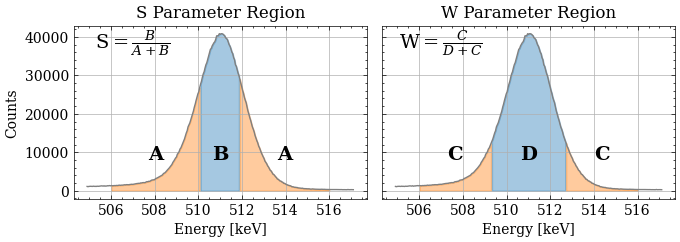}
\caption{Illustration of the definition of the line-shape parameters S and W used in positron annihilation spectroscopy (PAS). Note that the exact energy boundaries for calculating the S parameter may vary depending on the material system and experimental setup. The reason for that is the the resolution of the detector vary and thus the line-shape parameters are not comparable between systems}
\label{lineshape_parameters} 
\end{figure}

Characterization of the DB peak is traditionally done using two parameters: S (Shape) and W (Wings). These line-shape parameters reflect the fraction of annihilation events with low- and high-momentum electrons, respectively. Their definitions are illustrated in Figure~\ref{lineshape_parameters} where the boundaries of the various areas shown in the figures vary from system to system, as explained in the description of the figure. However, typical windows of the integration domains can range from 510.2 to 511.8 keV for region B and from 507.8 to 509.3 keV and from 512.7 to 514.8 keV for region C \cite{Selim2021}. 
Positrons trapped in defects tend to interact more strongly with conductance and valence electrons than with the core electrons. This is because the conductance electrons have small binding energy and their momentum distribution is wider than that of the core electrons. It results in smaller Doppler shifts compared to core electrons. Consequently, the S parameter increases with defect density, while the W parameter decreases. Note that this simplified picture does not account for the specific affinity of positrons for certain types of defects, which may in some cases repel them and thus exclude them from contributing to the annihilation signal.
Typically, the main interest in the line-shaped parameters lies not in their absolute values but in how these values vary with the positron implantation energy, which provides depth-resolved information about defect types and the electronic structure.

\clearpage

\section{Calculation Methods}\label{Methods}
 \texttt{PyPAS} modules are designed to be modular and make use of standard Python libraries for data analysis. This modularity allows researchers to conduct in-depth investigations and apply the tools flexibly according to their specific needs. It is stressed that the various modules of \texttt{PyPAS} can be used independently of each other. This is important because, for example, the diffusion length can be analyzed without the analyzing the spectroscopy data itself and vice versa. The following sections present the essential parts of the \texttt{PyPAS} package , with each section dedicated to a different module. 

\subsection{Annihilation line spectroscopy}\label{peak_analysis}
The \texttt{PASdb} object holds the annihilation peak on which it is able to apply various of standard spectroscopy operations such as background subtraction, summation, center and fwhm estimation, and so on. The traditional spectral analysis is primarily handled via the \texttt{PySpectrum} package, whose description is outside the scope of this work and is available at: \url{https://github.com/achiyaAmrusi/pySpectrum}. \texttt{PASdb} is also able to extract the line-shape parameters from the annihilation peak.  In contrast to standard gamma spectroscopy, where peaks are typically approximated using analytical fits (e.g., Gaussian functions), DBS requires a more direct characterization method. This is because the annihilation peak in DBS is not a simple Gaussian, but rather a convolution of the electron momentum distribution projected along the detector axis with the detector's energy resolution. For this reason, \texttt{PyPAS} uses a direct summation method for the extraction of the S and W parameters. In this process, the background—assumed to follow an error function shape—is subtracted from the spectrum. The remaining peak is then integrated over user-defined energy intervals to obtain the S and W values.
The \texttt{PAScdb} object is built from a list of photon energies corresponding to valid CDB events. From this data, \texttt{PAScdb} constructs the 2D CDB distribution and extracts the DB and energy resolution spectra (Figure~\ref{fig:cdb_2d_example}). The 2D CDB distribution is stored as \texttt{xarray.DataArray} objects, allowing convenient manipulation and visualization. The DB and energy resolution are extracted as \texttt{PASdb} and \texttt{Peak} respectively where the \texttt{Peak} object is from the \texttt{PySpectrum} package. 
\texttt{PyPAS} also includes a basic yet flexible module for filtering time-energy event lists from two detectors synchronized to the same clock. Due to the package modular structure, different filtering strategies can be applied independently while remaining within the \texttt{PyPAS} framework.
Within \texttt{PyPAS}, \texttt{PySpectrum} is used to load spectra, extract the DB peak using a signal-to-noise ratio method, compute peak properties, and perform background subtraction. It utilizes the uncertainties package for proper error propagation \cite{LEBIGOT}.

\subsection{Positron Implantation Profile}\label{positron_implantation}
The \texttt{PyPAS} positron implantation profile module includes a library of Makhov and Ghosh profile parameters, along with functions for generating implantation profiles based on them. The library contains published parameter sets for common elements and materials, primarily taken from the literature \cite{Dryzek2008,Ghosh1995}. For example, the implantation of the Ghosh profile in copper for several positron energies is shown in Figure \ref{fig:ghosh_profiles}.
The module also allows the generation of implantation profiles using custom parameters, even if they are not included in the built-in library. Additionally, importing external profiles is straightforward, a \texttt{PyPAS} represents implantation profiles using the standard \texttt{xarray.DataArray} framework, enabling easy integration and manipulation.
 The module includes a dedicated function for handling multilayer samples, where the total implantation distribution is built by stitching together the cumulative distributions of each individual layer. This approach avoids discontinuities that would arise from simply concatenating PDFs. 
Although this method is physically motivated and useful for estimating implantation in multilayered geometries, it is an approximation and may not capture all scattering effects at interfaces. The accuracy of the method has not yet been verified and, for accurate modeling in complex or high-energy cases, Monte Carlo simulations are recommended.

\subsection{Positron Annihilation Profile Solver}\label{annhilation_profile}
\texttt{PyPAS} implements a finite difference method to solve the one-dimensional positron diffusion equation (equation~\ref{eq:diffusion_equation}) using boundary conditions suitable for a finite-sized sample (equation~\ref{eq:finite_boundry_condition}).
\begin{equation}
D_+ \frac{dc(z)}{dz}\Big|_{z=z_{edge}} - \frac{c(z_{edge})}{L_+} = 0
\label{eq:finite_boundry_condition}
\end{equation}
The spatial domain is discretized into a mesh of points $z_i$, where $i\in[N]$, with each point corresponding to a specific depth within the sample. The structural and transport properties of the sample such as layer thicknesses, annihilation rates, diffusion coefficients, and mobilities are encapsulated in a \texttt{Sample} object, which is general enough to fit for lifetime analysis too. Additionally, the electric field profile in the sample is provided as an \texttt{xrarray.DataArray}.
Using these inputs, \texttt{PyPAS} constructs a tridiagonal matrix $M$ representing the discretized form of equation~\ref{eq:diffusion_equation}.
The discretizations results in a system of algebraic equations in equation~\ref{eq:discrete_diffusion_equation}). 

\begin{equation}
D_{i+\frac{1}{2}} \frac{c_{i+1} - c_{i}}{\Delta z^2} 
- D_{i-\frac{1}{2}} \frac{c_{i} - c_{i-1}}{\Delta z^2} 
- \frac{v_{i+1} c_{i+1} - v_{i-1} c_{i-1}}{2\Delta z} 
- \left( \lambda_b + \sum_d \lambda_d \right)_i \cdot c_i 
= - I_i
\label{eq:discrete_diffusion_equation}
\end{equation}

The boundary conditions are incorporated using the ghost point method, which allows for the application of derivative boundary conditions by introducing fictitious points outside the physical domain.  These boundary conditions are presented in equation~\ref{eq:discrete_diffusion_boundary_condition}. Notably, \texttt{PyPAS} evaluates physical quantities at the centers of mesh cells rather than at their edges, consistent with the central difference scheme commonly used in finite difference methods. Since the simulation domain is finite, radiative boundary conditions are applied at both ends to approximate open-system behavior.

\begin{equation}
\begin{cases}
\left( 2\cdot\frac{D_{0}}{\Delta z^2}\right) c_1 
- \left( \frac{D_{1/2}}{\Delta z^2} + \frac{D_{-1/2}}{\Delta z^2} + \lambda_{\text{eff}} - \frac{2\Delta z}{L_a}\left( \frac{D_{-1/2}}{\Delta z^2} + \frac{v_{-1}}{2\Delta z} \right) \right) c_0 
=  -I_0, \\
-\left( 2\frac{D_{N-\frac{1}{2}}}{\Delta z^2}\right) c_{N-1} 
- \left( \frac{D_{N+\frac{1}{2}}}{\Delta z^2} + \frac{D_{N-\frac{1}{2}}}{\Delta z^2} + \lambda_{\text{eff}} 
+ 2\frac{\Delta z}{L_+}\left( \frac{D_{N+\frac{1}{2}}}{\Delta z^2} - \frac{v_{N+1}}{2\Delta z} \right) \right) c_N 
= -I_N
\end{cases}
\label{eq:discrete_diffusion_boundary_condition}
\end{equation}

Once the matrix is constructed, \texttt{PyPAS} solves the system of the linear equations in equation \ref{eq:equation_set} using standard SciPy methods.

\begin{equation}
M_{ij} c(z_j) = -Im_j,
\label{eq:equation_set}
\end{equation}

where \( c(z_j) \) represents the positron annihilation rate in each spatial bin, and \( Im_j \) is the positron implantation profile at the same location. .

\begin{figure}
\centering
\includegraphics[width=0.95\linewidth]{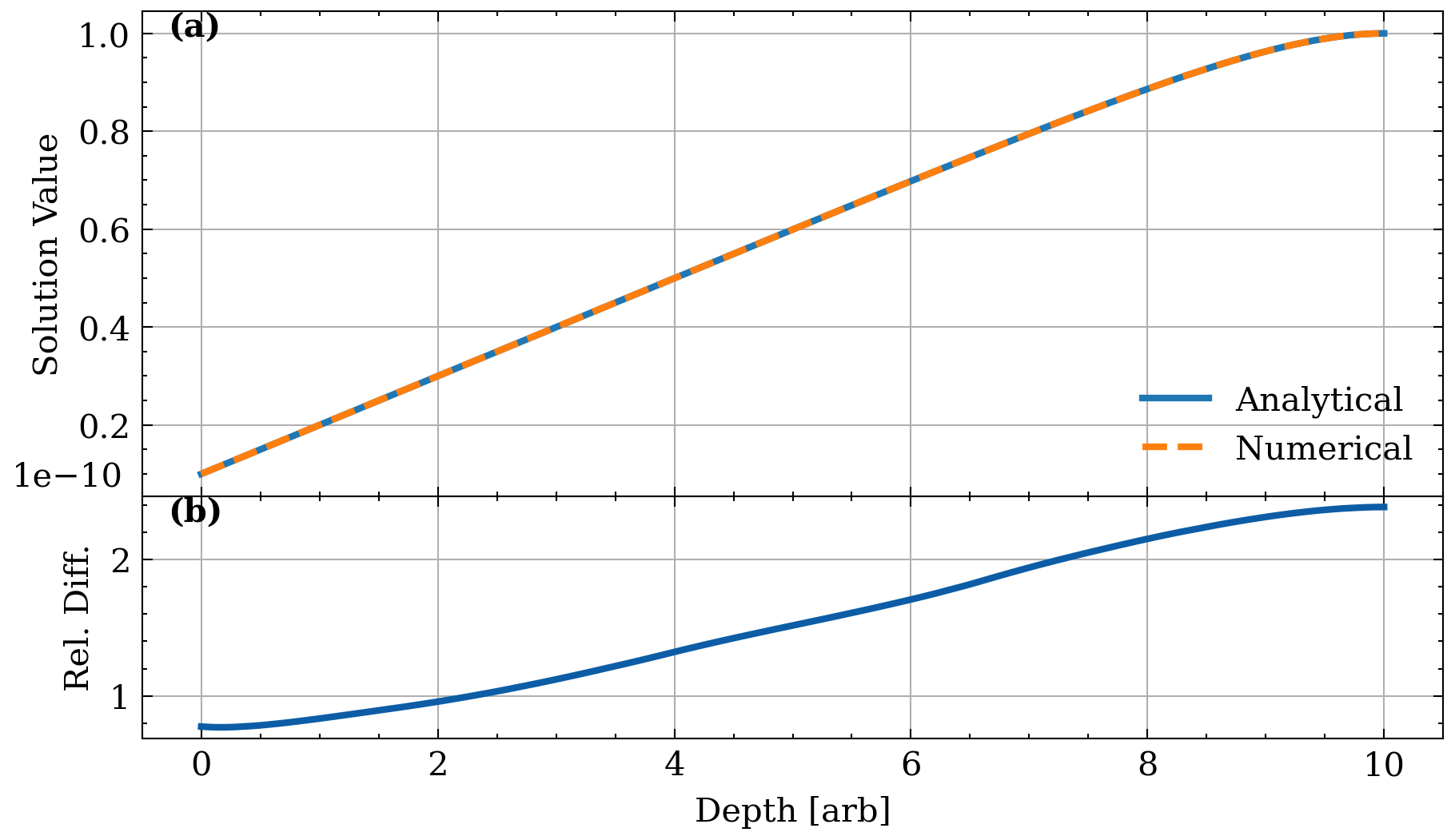}
\caption{Benchmark comparison between the numerical \texttt{PyPAS} solver and the analytical solution of the positron diffusion equation under idealized conditions. Both solutions assume a constant electric field, constant diffusion coefficient, uniform absorption probability, and a fixed positron implantation profile. For simplicity, the bulk annihilation rate was set to zero. The top panel shows excellent agreement between the numerical (dashed red) and analytical (solid blue) solutions across the spatial domain. The bottom panel displays the point-wise relative difference between the two, demonstrating that the \texttt{PyPAS} solver achieves high numerical accuracy down to the level of $1\mathrm{e}{-10}$  in terms of relative error.}
\label{fig:validation_analytical_test} 
\end{figure}

To validate the \texttt{PyPAS} solver, its numerical solution are compared to the analytical solution of the one-dimensional positron diffusion equation under conditions where the diffusion coefficient, mobility, annihilation rate, and electric field are constants. Figure~\ref{fig:validation_analytical_test} shows the relative difference between the \texttt{PyPAS} numerical result and the analytical solution, demonstrating the accuracy of the solver under these controlled conditions.

Two verification cases were also tested:
\begin{enumerate}
    \item Comparison between the \texttt{PyPAS} solver and \texttt{SciPy}'s ODE solver for the annihilation profile of a 10\,keV positron flux implanted into a $Si$/$SiO_2$ sample constructed of 200 nm oxide layer.
    
    \item Comparison of the percentage of positron annihilation rate in the bulk between the \texttt{PyPAS} numerical solution and the exact solution for a homogeneous sample with a given implantation profile, without an electric field, across various diffusion and absorption length models.
\end{enumerate}

\begin{figure}
\centering
\includegraphics[width=0.95\linewidth]{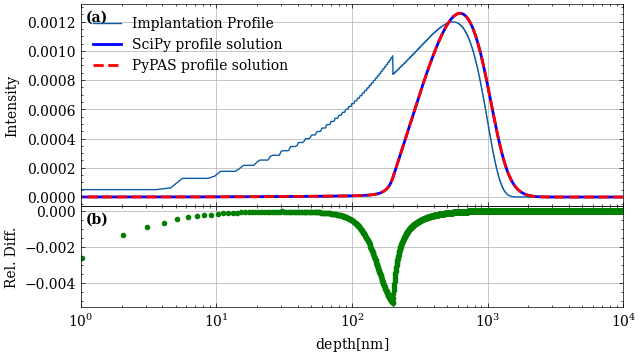}
\caption{ \textbf{Top}:Comparison of positron annihilation profiles in a simulated Si wafer. The Ghosh thermal implantation profile is shown in light blue. Annihilation profiles calculated using the  \texttt{SciPy} BVP solver (dark blue) and the \texttt{PyPAS} solver (red) are overlaid. Both solvers yield nearly identical results. \textbf{Bottom}: Relative difference between the \texttt{PyPAS} and analytical solutions for the fraction of 10\,keV positrons annihilated at the surface. The maximum deviation is less than 0.6\%. Simulation parameters are detailed in table~\ref{tab:Si_wafer_sample}. The finite difference algorithm was applied with 100{,}000 spatial cells compared with 10000 on SciPy.}
\label{fig:pypas_and_scipy_si} 
\end{figure}

\begin{figure}
\centering
\includegraphics[width=0.95\linewidth]{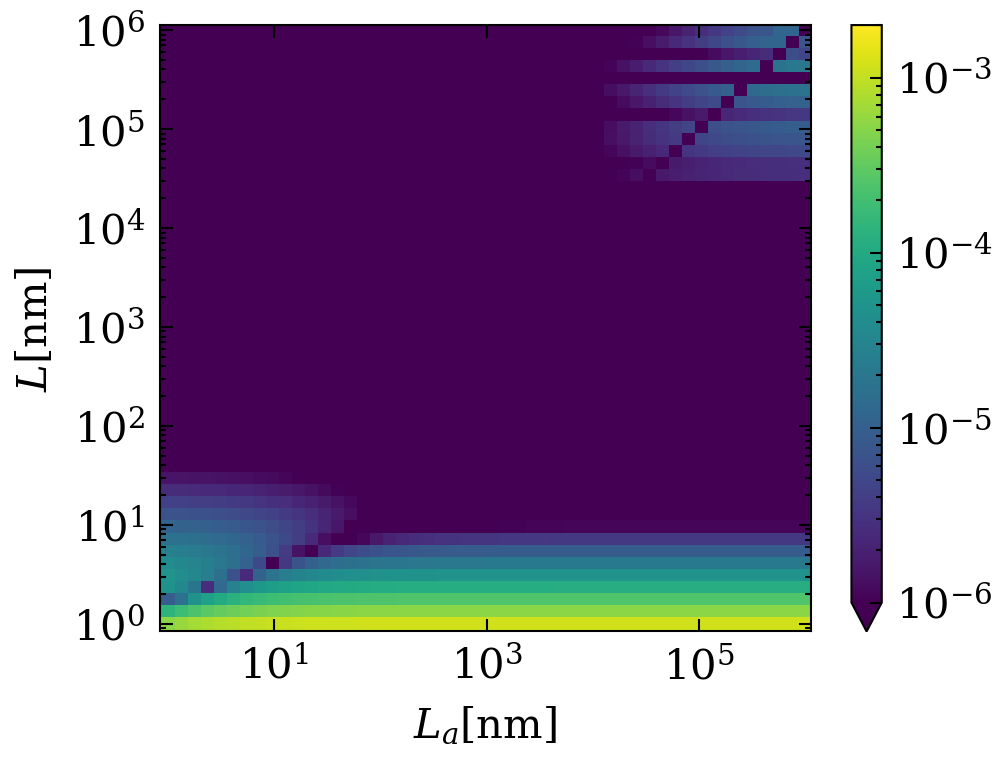}
\caption{Relative deviation in the bulk positron annihilation fraction between the  \texttt{PyPAS} and analytical solutions for the fraction of 10\,keV positrons annihilated at the surface. The maximum deviation is less than 0.12\%. Accuracy of the \texttt{PyPAS} solution decreases when the positron diffusion length approaches the mesh resolution or the sample thickness. Simulation parameters are provided in table~\ref{tab:Cu_Sample_and_Implantation}. The finite difference algorithm was computed using 100{,}000 spatial cells.}
\label{fig:pypas_vs_analytical} 
\end{figure}

In both cases, the \texttt{PyPAS} solver demonstrated strong agreement with reference solutions. Notably, in the first case, minor deviations were observed at the $Si$/$SiO_2$ interface, likely due to convergence challenges faced by the \texttt{SciPy} solver in regions with abrupt material property changes. In general, the relative error remained below 1\% throughout the domain, underscoring the robustness and precision of the \texttt{PyPAS} solver. The results are presented in figure~\ref{fig:pypas_and_scipy_si} and figure~\ref{fig:pypas_vs_analytical}.

\begin{table}[h]
    \centering
    \renewcommand{\arraystretch}{1.3}
    \begin{tabular}{|c|c|c|c|}
        \hline
        \textbf{Layer} & \textbf{Material} & \textbf{Diffusion Length} \(L_+\) [nm]& \textbf{Width} [nm]\\
        \hline
        Layer 1 & \(SiO_2\) & \( 20 \)& \( 2 \)\\
        \hline
        Layer 2 & \(Si\) & \( 220\)& \( 10^4\)\\
        \hline
        \multicolumn{4}{|c|}{\textbf{Absorption Length on Surface} \( L_a \) (nm)} \\
        \hline
        \multicolumn{4}{|c|}{\( 0.1 \)} \\
        \hline
        \multicolumn{4}{|c|}{\textbf{Ghosh Implantation Parameters}} \\
        \hline
        \multicolumn{2}{|c|}{\textbf{Parameter}} & \textbf{SiO\(_2\)} & \textbf{Si} \\
        \hline
        \multicolumn{2}{|c|}{Density (g/cm\(^3\))} & 2.65 & 2.329 \\
        \multicolumn{2}{|c|}{\(m\)} & 4.47 & 4.05 \\
        \multicolumn{2}{|c|}{\(l\)} & 0.61 & 0.481 \\
        \multicolumn{2}{|c|}{\(c_{lm}\)} & 1.622 & 1.671 \\
        \multicolumn{2}{|c|}{\(N_{lm}\)} & 1.107 & 0.995 \\
        \multicolumn{2}{|c|}{\(n\)} & 1.772 & 1.625 \\
        \multicolumn{2}{|c|}{\(B\) (nm/$keV^n$)} & 8.36& 13.6\\
        \hline
    \end{tabular}
    \caption{Description of the Si wafer sample layers, absorption properties, and positron implantation parameters for SiO\(_2\) and Si.}
    \label{tab:Si_wafer_sample}
\end{table}

\begin{table}[h]
    \centering
    \renewcommand{\arraystretch}{1.3}
    \begin{tabular}{|c|c|c|c|}
        \hline
        \textbf{Layer} & \textbf{Material} & \textbf{Diffusion Length} \(L_+\) [nm]& \textbf{Width} [nm]\\
        \hline
        Layer 1 & - & \(1 - 10^6\)& \( 10^4 \)\\
        \hline
        \multicolumn{4}{|c|}{\textbf{Absorption Length on Surface} \( L_a \) (nm)} \\
        \hline
        \multicolumn{4}{|c|}{\(1 - 10^{6}\)} \\
        \hline
        \multicolumn{4}{|c|}{\textbf{Ghosh Implantation Parameters}} \\
        \hline
        \multicolumn{2}{|c|}{\textbf{Parameter}} & \multicolumn{2}{c|}{\textbf{Cu}} \\
        \hline
        \multicolumn{2}{|c|}{Density (g/cm\(^3\))} & \multicolumn{2}{c|}{8.96} \\
        \multicolumn{2}{|c|}{\(m\)} & \multicolumn{2}{c|}{3.02} \\
        \multicolumn{2}{|c|}{\(l\)} & \multicolumn{2}{c|}{0.374} \\
        \multicolumn{2}{|c|}{\(c_{lm}\)} & \multicolumn{2}{c|}{1.653} \\
        \multicolumn{2}{|c|}{\(N_{lm}\)} & \multicolumn{2}{c|}{0.929} \\
        \multicolumn{2}{|c|}{\(n\)} & \multicolumn{2}{c|}{1.719} \\
        \multicolumn{2}{|c|}{\(B\) (nm/$\text{keV}^n
$)} & \multicolumn{2}{c|}{2.544} \\
        \hline
    \end{tabular}
    \caption{Description of the Cu sample layers, absorption properties, and positron implantation parameters.}
    \label{tab:Cu_Sample_and_Implantation}
\end{table}

\subsection{Diffusion Length Optimization}\label{diffusion_length_optimization}
Variable-energy Doppler broadening (VEDB) measurements are commonly used to extract the positron diffusion length by fitting experimental data to theoretical models, as implemented in tools such as \texttt{VEPFIT} and others~\cite{Veen1991,Dryzek2021,Saleh1999}. \texttt{PyPAS} includes several \texttt{SciPy}-based fitting schemes for layered sample models. However, unlike traditional software, \texttt{PyPAS} is not a black-box tool: it offers flexible, transparent fitting capabilities that allow deeper and customized analysis.

For example, a recent study ~\cite{Or2024} reported VEDB measurements of an annealed copper sample (annealed at 500\,$^\circ$C for 10 hours), where the diffusion length was estimated at 126\,nm using a two-layer model in \texttt{VEPFIT}. In that model, the first layer is thin (40\,nm) and the second layer represents the bulk material. In this work, it is demonstrated using \texttt{PyPAS} that the proposed two-layer model is not necessarily preferable to a simpler, one-layer model.

\begin{figure}
\centering
\includegraphics[width=0.95\linewidth]{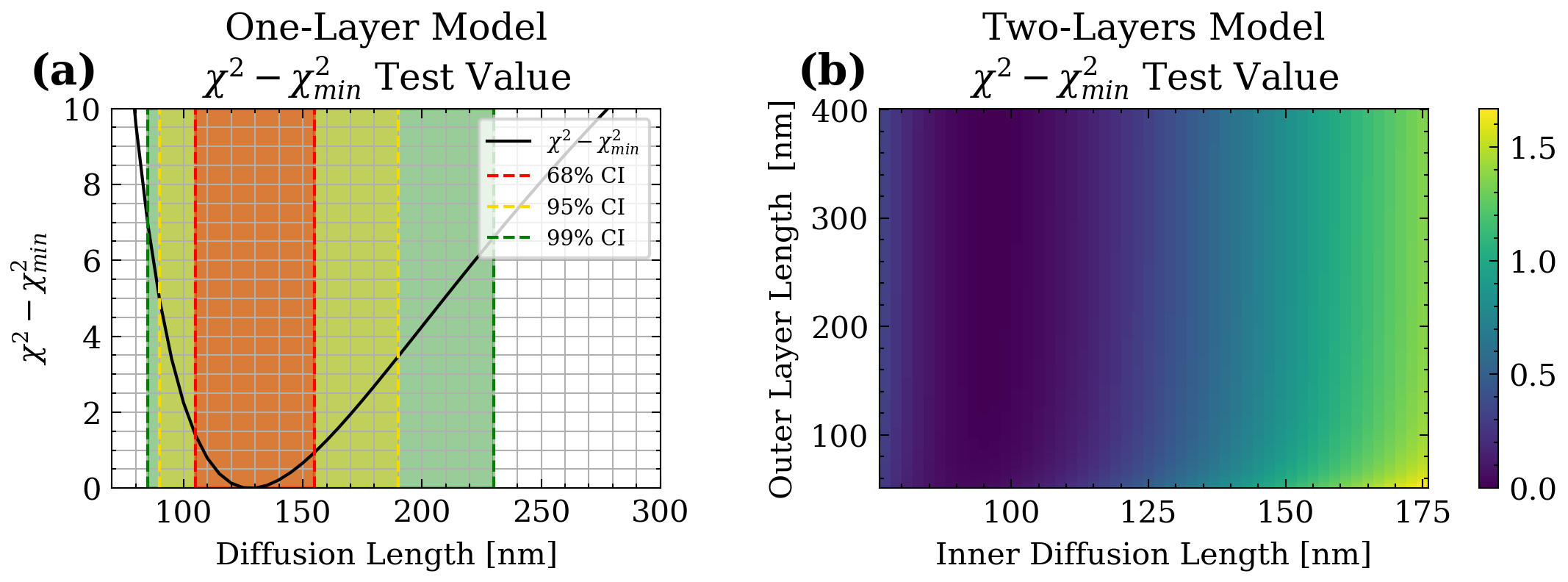}
\caption{ \textbf{Top}:Comparison of $\chi^2$ values for positron annihilation profile fits in one- and two-layer diffusion models. \textbf{(a)} The $\chi^2$ test values as a function of the diffusion length for the one-layer model. Shaded regions indicate the 68\%, 95\%, and 99\% confidence intervals \textbf{(b)} A 2D figure of $\chi^2$ values over first-layer and second-layer diffusion lengths in the two-layer model. The color scale reflects the $\chi^2$ values. It is clear that the chi-squared values do not change drastically and the fitting of the data at extreme values of the first layer diffusion length indicate an over-fit}
\label{fig:chi_square_test} 
\end{figure}

This re-analysis is motivated in part by findings from previous work~\cite{Amrusi2024}, which showed that the SPOT-IL beam exhibits increased noise at higher positron implantation energies --- likely due to beam movement. That study used \texttt{PyPAS} to identify this noise, and based on its findings,  restrict the analysis to implantation energies up to 8.5\,keV, where the data points are more reliable. Despite this limitation, the analysis remains sufficient: the mean implantation depth for 8\,keV positrons in copper is approximately 130 nm --- deeper than the proposed thin surface layer. Therefore, the measured data still contain information about the near-surface region and can be used to assess the existence of such a layer.
In this example, the chi-squared values for one- and two-layer models of diffusion length are computed and compared with respect to diffusion length. Figure~\ref{fig:chi_square_test} shows the chi-squared test values and the confidence interval 69\%, 95\% and 99\% for the one-layer model. Figure~\ref{fig:chi_square_test} also presents the corresponding chi-square test values for the two-layer model. Although the two-layer model shows a region of low chi-square values, the bulk diffusion length of the minimum chi-squared domain overlaps with the confidence interval obtained for the one-layer model. The region of minimal chi-squared values of the first model diffusion length is not physical due to the fitting in the extreme value of up to 400 nm. Moreover, the improvement in the minimum chi-squared value is marginal -- approximately 0.4 -- suggesting a possible overfit. To compare the models' performance, the Akaike Information Criterion (AIC) is calculated at the minimum chi-squared value for each model. The resulting AIC values are 18.6 for the one-layer model and 20.3 for the two-layer model. This suggests that the one-layer model is approximately 2.3 times more likely than the two layer model. 
This example illustrates the power and flexibility of \texttt{PyPAS}, enabled by its modular design and integration within the Python ecosystem. Importantly, the ability to formulate and test competing hypotheses, such as comparing structural models of the sample, is essential specifically in variable energy analysis, where the depth-dependent sensitivity of the measurement makes model selection critical. The open-source and transparent nature of \texttt{PyPAS} empowers researchers to explore alternative scenarios, validate assumptions, and perform reproducible analysis beyond the constraints of traditional black-box fitting tools. 

\clearpage

\section{Conclusions}\label{Conclusions}

\texttt{PyPAS} is an open-source Python package for positron annihilation spectroscopy (PAS), designed to handle Doppler broadening and two-dimensional coincidence Doppler broadening spectra. It extracts line-shape parameters, calculates positron annihilation profiles, and supports variable-energy Doppler broadening analysis. The package builds upon the PySpectrum library, which provides tools for spectrum and peak handling, including the extraction of features such as FWHM, mean energy, and integrated counts.

\texttt{PyPAS} enables comprehensive positron analysis within a modern Python environment, streamlining the workflow from raw spectrum extraction to diffusion length optimization. It includes a set of well-documented examples—covering both synthetic and experimental data—that are publicly available on GitHub.
In this work, the accuracy and reliability of \texttt{PyPAS} were demonstrated through validation against analytical solutions and verification using the  boundary value problem solver of \texttt{SciPy}. The framework supports the integration of modern analysis techniques, such as Bayesian inference, and promotes a modular, transparent approach that offers greater flexibility compared to traditional black-box methods. This was exemplified through hypothesis testing in a VEDB analysis of an annealed Cu sample, where \texttt{PyPAS} enabled a critical reassessment of a previously proposed two-layer model.
Future extensions of \texttt{PyPAS} are planned to include positron lifetime analysis, with the goal of providing a unified, Python-based toolkit for all standard PAS techniques.

\section*{CRediT authorship contribution statement}
A. Y. Amrusi: Investigation, Software. \\
S. May-Tal Beck: Supervision, Conceptualization. \\
H. Steinberg: Supervision, Conceptualization. \\
G. Ron: Supervision, Conceptualization.

\section*{Declaration of competing interest}
The authors declare that they have no known competing financial interests or personal relationships that could have appeared to influence the work reported in this paper.

\section*{Declaration of generative AI and AI-assisted technologies in the writing process}
During the preparation of this work the authors used ChatGPT to improve the readability of the code, documentation, and manuscript. After using this tool, the authors reviewed and edited the content as needed and take full responsibility for the content of the published article.

\section*{Data availability}
Data are available on GitHub at \url{https://github.com/achiyaAmrusi/pyPAS} in the examples directory.

\section*{Acknowledgments}
This project was supported by the Ministry of Economy, under the KAMIN program and by the Pazy Research Foundation Grant No. 519/2023, and Pazy equipment grant of the Israel Atomic Energy Com mission.


\begin{thebibliography}{11}
\expandafter\ifx\csname natexlab\endcsname\relax\def\natexlab#1{#1}\fi
\providecommand{\url}[1]{\texttt{#1}}
\providecommand{\href}[2]{#2}
\providecommand{\path}[1]{#1}
\providecommand{\DOIprefix}{doi:}
\providecommand{\ArXivprefix}{arXiv:}
\providecommand{\URLprefix}{URL: }
\providecommand{\Pubmedprefix}{pmid:}
\providecommand{\doi}[1]{\href{http://dx.doi.org/#1}{\path{#1}}}
\providecommand{\Pubmed}[1]{\href{pmid:#1}{\path{#1}}}
\providecommand{\bibinfo}[2]{#2}
\ifx\xfnm\relax \def\xfnm[#1]{\unskip,\space#1}\fi
\bibitem[{Tuomisto and Makkonen(2013)}]{Tuomisto2013}
\bibinfo{author}{F.~Tuomisto}, \bibinfo{author}{I.~Makkonen},
\newblock \bibinfo{title}{Defect identification in semiconductors with positron annihilation: Experiment and theory},
\newblock \bibinfo{journal}{Rev. Mod. Phys.} \bibinfo{volume}{85} (\bibinfo{year}{2013}) \bibinfo{pages}{1583--1631}. \URLprefix \url{https://link.aps.org/doi/10.1103/RevModPhys.85.1583}. \DOIprefix\doi{10.1103/RevModPhys.85.1583}.
\bibitem[{Selim(2021)}]{Selim2021}
\bibinfo{author}{F.~Selim},
\newblock \bibinfo{title}{Positron annihilation spectroscopy of defects in nuclear and irradiated materials- a review},
\newblock \bibinfo{journal}{Materials Characterization} \bibinfo{volume}{174} (\bibinfo{year}{2021}) \bibinfo{pages}{110952}. \URLprefix \url{https://www.sciencedirect.com/science/article/pii/S1044580321000826}. \DOIprefix\doi{https://doi.org/10.1016/j.matchar.2021.110952}.
\bibitem[{Veen et~al.(1991)Veen, Schut, Vries, Hakvoort, and Ijpma}]{Veen1991}
\bibinfo{author}{A.~v. Veen}, \bibinfo{author}{H.~Schut}, \bibinfo{author}{J.~d. Vries}, \bibinfo{author}{R.~A. Hakvoort}, \bibinfo{author}{M.~R. Ijpma},
\newblock \bibinfo{title}{{Analysis of positron profiling data by means of ‘‘VEPFIT’’}},
\newblock \bibinfo{journal}{AIP Conference Proceedings} \bibinfo{volume}{218} (\bibinfo{year}{1991}) \bibinfo{pages}{171--198}. \URLprefix \url{https://doi.org/10.1063/1.40182}. \DOIprefix\doi{10.1063/1.40182}. \href{http://arxiv.org/abs/https://pubs.aip.org/aip/acp/article-pdf/218/1/171/12126267/171\_1\_online.pdf}{{\tt arXiv:https://pubs.aip.org/aip/acp/article-pdf/218/1/171/12126267/171\_1\_online.pdf}}.
\bibitem[{Saleh et~al.(1999)Saleh, Taylor, and Rice-Evans}]{Saleh1999}
\bibinfo{author}{A.~Saleh}, \bibinfo{author}{J.~Taylor}, \bibinfo{author}{P.~Rice-Evans},
\newblock \bibinfo{title}{The royprof program for analyzing positron profiling data obtained from variable energy beams},
\newblock \bibinfo{journal}{Applied Surface Science} \bibinfo{volume}{149} (\bibinfo{year}{1999}) \bibinfo{pages}{87--96}. \URLprefix \url{https://www.sciencedirect.com/science/article/pii/S0169433299001798}. \DOIprefix\doi{https://doi.org/10.1016/S0169-4332(99)00179-8}.
\bibitem[{Dryzek(2021)}]{Dryzek2021}
\bibinfo{author}{J.~Dryzek},
\newblock \bibinfo{title}{Analysis of positron profiling data using e+dsc computer code},
\newblock \bibinfo{journal}{Computer Physics Communications} \bibinfo{volume}{264} (\bibinfo{year}{2021}) \bibinfo{pages}{107937}. \URLprefix \url{https://www.sciencedirect.com/science/article/pii/S0010465521000667}. \DOIprefix\doi{https://doi.org/10.1016/j.cpc.2021.107937}.
\bibitem[{Ghosh(1995)}]{Ghosh1995}
\bibinfo{author}{V.~Ghosh},
\newblock \bibinfo{title}{Positron implantation profiles in elemental and multilayer systems},
\newblock \bibinfo{journal}{Applied Surface Science} \bibinfo{volume}{85} (\bibinfo{year}{1995}) \bibinfo{pages}{187--195}. \URLprefix \url{https://www.sciencedirect.com/science/article/pii/0169433294003319}. \DOIprefix\doi{https://doi.org/10.1016/0169-4332(94)00331-9}, \bibinfo{note}{proceedings of the Sixth International Workshop on Slow-Positron Beam Techniques for Solids and Surfaces}.
\bibitem[{Dryzek and Horodek(2008)}]{Dryzek2008}
\bibinfo{author}{J.~Dryzek}, \bibinfo{author}{P.~Horodek},
\newblock \bibinfo{title}{Geant4 simulation of slow positron beam implantation profiles},
\newblock \bibinfo{journal}{Nuclear Instruments and Methods in Physics Research Section B: Beam Interactions with Materials and Atoms} \bibinfo{volume}{266} (\bibinfo{year}{2008}) \bibinfo{pages}{4000--4009}. \URLprefix \url{https://www.sciencedirect.com/science/article/pii/S0168583X08008926}. \DOIprefix\doi{https://doi.org/10.1016/j.nimb.2008.06.033}.
\bibitem[{Butterling et~al.(2011)Butterling, Anwand, Cowan, Hartmann, Jungmann, Krause-Rehberg, Krille, and Wagner}]{BUTTERLING20112623}
\bibinfo{author}{M.~Butterling}, \bibinfo{author}{W.~Anwand}, \bibinfo{author}{T.~E. Cowan}, \bibinfo{author}{A.~Hartmann}, \bibinfo{author}{M.~Jungmann}, \bibinfo{author}{R.~Krause-Rehberg}, \bibinfo{author}{A.~Krille}, \bibinfo{author}{A.~Wagner},
\newblock \bibinfo{title}{Gamma-induced positron spectroscopy (gips) at a superconducting electron linear accelerator},
\newblock \bibinfo{journal}{Nuclear Instruments and Methods in Physics Research Section B: Beam Interactions with Materials and Atoms} \bibinfo{volume}{269} (\bibinfo{year}{2011}) \bibinfo{pages}{2623--2629}. \URLprefix \url{https://www.sciencedirect.com/science/article/pii/S0168583X11006513}. \DOIprefix\doi{https://doi.org/10.1016/j.nimb.2011.06.023}.
\bibitem[{Or et~al.(2024)Or, Chelladurai, Cohen, Amrosi, Cohen, Sabo-Napadensky, Gordon, Cohen, Presler, Cohen, Piasetzky, Steinberg, Beck, and Ron}]{Or2024}
\bibinfo{author}{P.~Or}, \bibinfo{author}{L.~Chelladurai}, \bibinfo{author}{D.~Cohen}, \bibinfo{author}{A.~Amrosi}, \bibinfo{author}{T.~Cohen}, \bibinfo{author}{I.~Sabo-Napadensky}, \bibinfo{author}{E.~Gordon}, \bibinfo{author}{S.~Cohen}, \bibinfo{author}{O.~Presler}, \bibinfo{author}{E.~Cohen}, \bibinfo{author}{E.~Piasetzky}, \bibinfo{author}{H.~Steinberg}, \bibinfo{author}{S.~M.-T. Beck}, \bibinfo{author}{G.~Ron},
\newblock \bibinfo{title}{The spot-il positron beam construction and its use for doppler broadening measurement of annealed cu},
\newblock \bibinfo{journal}{Nuclear Instruments and Methods in Physics Research Section B: Beam Interactions with Materials and Atoms} \bibinfo{volume}{546} (\bibinfo{year}{2024}) \bibinfo{pages}{165174}. \URLprefix \url{https://www.sciencedirect.com/science/article/pii/S0168583X23004159}. \DOIprefix\doi{https://doi.org/10.1016/j.nimb.2023.165174}.
\bibitem[{Lebigot(nd)}]{LEBIGOT}
\bibinfo{author}{E.~O. Lebigot}, \bibinfo{title}{Uncertainties: a python package for calculations with uncertainties}, \bibinfo{year}{n.d.} \URLprefix \url{http://pythonhosted.org/uncertainties/}.
\bibitem[{Amrusi(2024)}]{Amrusi2024}
\bibinfo{author}{A.~Y. Amrusi}, \bibinfo{title}{Reliability study of the slow positron beam in SPOT-IL}, \bibinfo{type}{Master's thesis}, The Racah Institute of Physics, Hebrew University of Jerusalem, \bibinfo{address}{Jerusalem, Israel}, \bibinfo{year}{2024}. \bibinfo{note}{Supervised by Prof. Guy Ron and Prof. Hadar Steinberg}.

\end{thebibliography}
\end{document}